\documentstyle[12pt,moriond,epsfig]{article}

\begin{document}
\heading{Spatial resolution bias in the mid-infrared Starburst/AGN classification}
\vspace{-3mm}
\author{O. Laurent $^{1}$, I.F. Mirabel $^{1,2}$, V. Charmandaris $^{3}$, P. Gallais $^{1}$, \\M. Sauvage $^{1}$, L. Vigroux $^{1}$, C.J. Cesarsky $^{1}$} 
{$^{1}$ CEA/DSM/DAPNIA Service d'Astrophysique, F-91191 Gif-sur-Yvette, France.} 
{$^{2}$ Instituto de Astronom\'\i a y F\'\i sica del Espacio. cc 67, suc 28. 1428 Buenos Aires, Argentina.}
{$^{3}$ Observatoire de Paris, DEMIRM, 61 Av. de l'Observatoire, F-75014 Paris, France.}

\vspace{-4mm}
\begin{moriondabstract}
We present the effects of limited spatial resolution to the observed
mid-infrared (MIR) spectrum of an active galactic nucleus (AGN)
surrounded by a disk with massive star forming regions. Using MIR
observations of the face-on nearby Seyfert 1 galaxy NGC\,6814, we vary
the observing aperture and examine the evolution of the observed
AGN/starburst fraction with our MIR diagnostic. We show that the
spatial resolution of ISOCAM is sufficient to disentangle AGN from
starburst features in nuclear regions of nearby galaxies
(D\,$<$\,50\,Mpc). However, with the exception of a few ultra-luminous
galaxies, dilution effects hide completely the AGN contribution in
more distant galaxies.
\end{moriondabstract}

\vspace{-5mm}
\section{Introduction}
\vspace{-2mm}
Based on ground-based MIR observations (\cite{Roche}, \cite{Dudley})
and more recently with ISO (\cite{Lutz}, \cite{Genzelb} and references
therein), considerable progress has been made in defining the fraction
of the AGN/starburst contribution to the MIR spectral energy
distribution (SED) of luminous infrared galaxies.  To further examine
the AGN and starburst connection, we have developed a new MIR
diagnostic diagram using ISOCAM observations from 5 to 16\,$\mu m$,
which allows us to quantify emission due to AGNs from that resulting
from star formation activity (see \cite{Laurent1} for details). Our
diagram is based on the fact that MIR spectra of late type galaxies
--- assuming that the stellar contribution to active/star forming
regions is negligible --- can be decomposed in three components
characteristic of emission from HII regions, photo-dissociation
regions (PDRs) and AGNs. Each component presents an unique
signature: 1) the hot continuum at short wavelengths (3-5\,$\mu m$) is
present in AGNs (\cite{Genzela}, \cite{Dudley}, \cite{Laurent1}), 2)
Unidentified Infrared Bands (UIBs) at 6.2, 7.7, 8.6, 11.3 and
12.7\,$\mu m$ are dominant in PDRs as well as in diffuse regions
(\cite{Verstraete}, \cite{Tran}, \cite{Mattila}), and 3) a strong
continuum due to emission from very small grains is detected in HII
regions (\cite{Verstraete}, \cite{Cesarsky}).  Due to the small
physical size of the region which is heated directly by an AGN, a good
spatial resolution is essential in order to probe the physics of the
dust in nuclear regions hosting a weak AGN  (i.e. where the MIR
emission from the whole galaxy is largely higher than that from the
AGN).

In order to examine how the spatial resolution of a telescope (in our
case ISOCAM) may bias the detection of the MIR emission from an AGN
and its surrounding star formation regions, we used NGC\,6814, a nearby
face-on Seyfert 1 galaxy located at 20\,Mpc
(H$_{0}$=75\,km$^{-1}$Mpc$^{-1}$)(\cite{Ulrich}), as a test case.  The
following section displays the effects of dilution in the MIR spectrum
when we observe galaxies at increasingly larger distances. In Section
3, we show in detail how the integration beam affects the MIR SEDs as
well as the evolution of the observed spectrum on our diagnostic
diagram. Finally, we comment on the expected results from future
instruments such as SIRTF and the NGST.

\vspace*{-5mm}
\section{Contamination of an AGN from circum-nuclear starbursts.}

\begin{figure}[!ht]
\hspace{-5mm}
\resizebox{17cm}{!}{\includegraphics{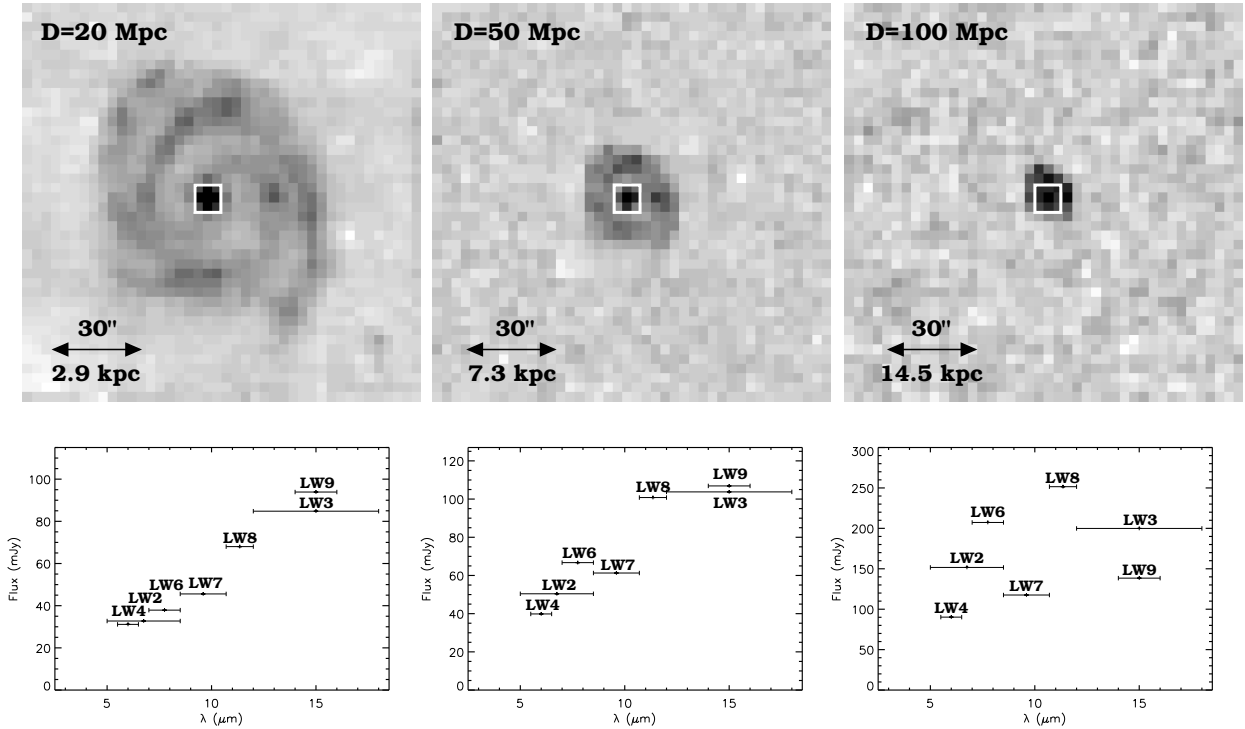}}
\caption{
\footnotesize{
Effects of the spatial resolution in a MIR AGN spectrum. As an example,
we use the Seyfert 1 galaxy NGC\,6814 actually located at
$\sim$20\,Mpc (H$_{0}$=75\,km$^{-1}$Mpc$^{-1}$). Left panel, we show
an image of NGC\,6814 in the 7-8.5$\mu m$ band tracing mainly the UIB
feature at 7.7$\mu m$. Middle panel, the galaxy is projected to a
distance of 50\,Mpc. Right panel, the projected distance is increased
to 100\,Mpc. For each distance, we calculate the flux inside a
constant ISOCAM aperture given by 3$\times$3 pixels
(i.e. 9''$\times$9'', see SEDs underneath images). We have also
applied an aperture correction to get spectra comparable to those in
Figure \ref{figure2}.}}
\label{figure1}
\end{figure}

The imaging capabilities of ISOCAM provide an angular resolution of
5''-\,8'' at 7-15\,$\mu m$ (250-400\,pc at 10\,Mpc). Even though in some
extreme cases of ultra-luminous galaxies (e.g. IRAS 19254-7245 in
\cite{Laurent2} and IRAS 23060+0505 in \cite{Genzela}), the AGN
contribution may completely dominate the total emission coming from a
galaxy, the MIR emission in most galaxies of our sample containing an
AGN is dominated by their disk emission. As a consequence, a good
spatial resolution is necessary to separate galactic nuclei
($\sim$1\,kpc in diameter) and to distinguish emission from an
AGN. Moreover, intrinsic absorption particularly in Seyfert 2
galaxies can hide the AGN even in the MIR making its detection
difficult.  We present in Figure \ref{figure1} images of NGC\,6814
projected at different distances as well as their corresponding
SEDs. Given a fixed aperture of 9'' (3\,pixels at 3''/pixel), we note
the increase of UIBs associated with star forming regions in the
galactic disk by observing distant galaxies. The AGN continuum is
still detected at 50\,Mpc, but it becomes negligible at 100\,Mpc.

\vspace{-5mm}
\section{MIR spectral classification: from an AGN to a galactic disk.}

\begin{figure}[!ht]
\hspace{0mm}
\resizebox{17cm}{!}{\includegraphics{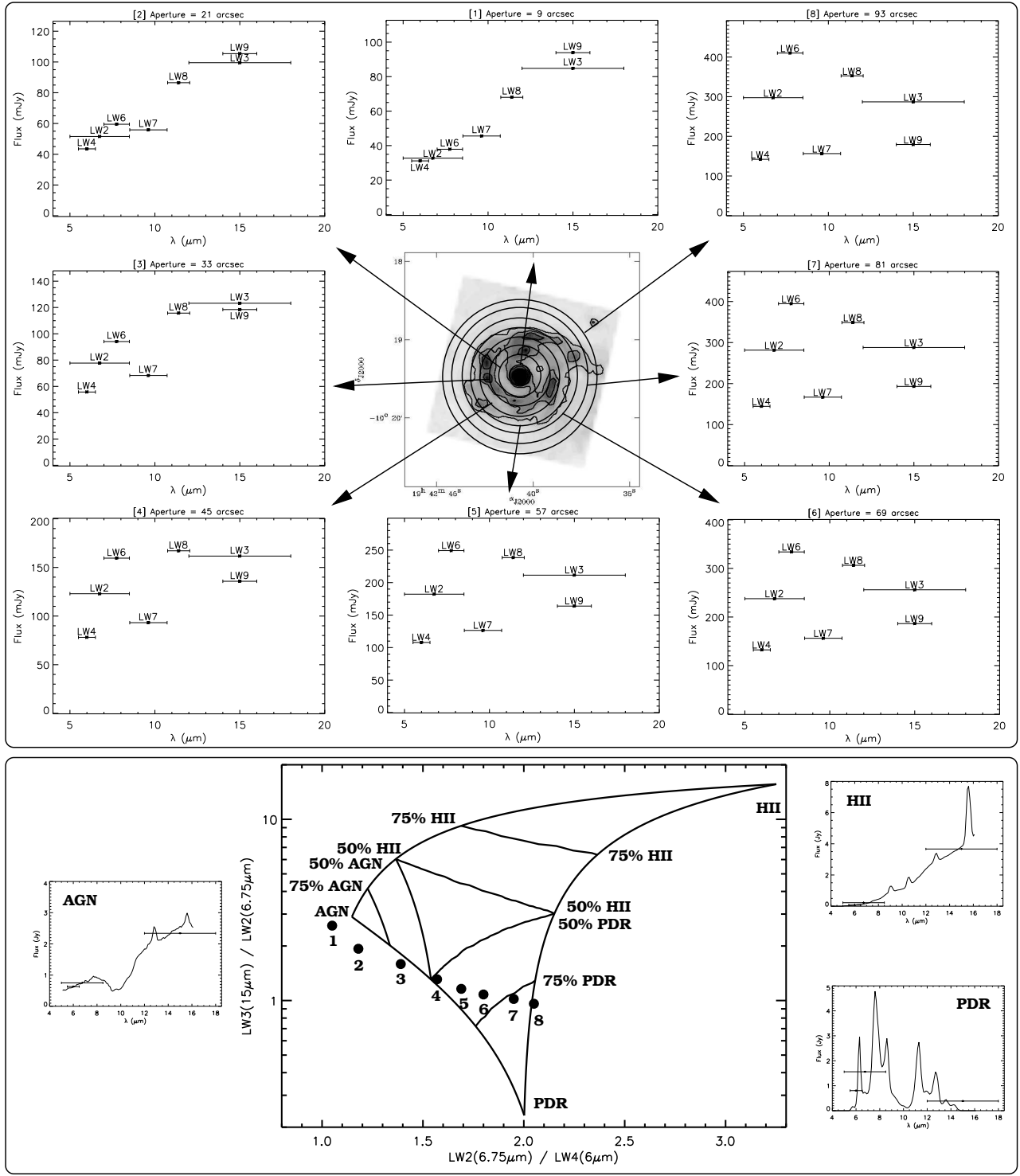}}
\caption{
\footnotesize{
Top panel: variation of the MIR emission as a function of the size of
the integrating aperture. We can follow (counter-clockwise) the
evolution of the shape of the SED for each aperture radius from the
AGN (Spectrum 1) to the total integrated galaxy (Spectrum 8). Note the
absence of UIBs and the strong rising continuum at short wavelengths
(5-8.5\,$\mu m$) in the central region (Spectrum 1) typical of AGN
emission. By increasing the aperture radius, the disk contribution
dominated by UIBs contaminates the AGN emission.  Bottom panel:
application of our MIR diagram ([5]), for all
SEDs.  Each point corresponds to a spectrum marked from 1 to 8. The
percentages indicate the fraction in the MIR (5-16\,$\mu m$) of each
template, in the observed spectra. The increasing disk contribution
changes our classification from ``AGN-like'' to the ``PDR-like'',
which is typical to galactic disks. The AGN dominates the MIR emission
only for small apertures smaller than 45'' in diameter (up to spectrum
4 with diameter\,=\,4.3\,kpc).}}
\label{figure2}
\end{figure}

Our new MIR diagnostic diagram classification method is based on the
UIB strength (LW2(5-8.5$\mu m$)/LW4(5.5-6.5$\mu m$)) and the slope of
the MIR continuum (LW3(12-18$\mu m$)/LW2(5-8.5$\mu m$))
(\cite{Laurent1}). In Figure \ref{figure2}, we present 8 SEDs
corresponding to aperture diameters from 9''(nucleus) to 93''(whole
galaxy) where we can observe the evolution from ``pure'' AGN spectrum
to ``pure'' PDR spectrum (i.e. the AGN continuum is not detected). We
can note the importance of the spatial resolution to find out MIR AGN
features in our diagnostic diagram.

\newpage

ISOCAM was the first instrument allowing us deep imaging covering the
total wavelength range between 5 and 16\,$\mu m$. Our diagnostic
method can successfully detect the presence of an AGN in
galaxies. However, it is limited to finding AGNs in nearby galaxies at
distances less than 50\,Mpc with the exception of some ultra-luminous
galaxies powered principally by AGN. For galaxies at z$>$0.1, the
redshift effect begins to play an important role since we look at
through fixed broad band filters and as a consequence the shape of our
diagnostic changes.

MIR observations IR-luminous galaxies at distances up to z=1, require
new infrared instruments such as SIRTF and the NGST who will provide
better spatial resolution to study the relationship between starburst
and AGN emission. Moreover, due to their higher sensitivity, one could
also compare our results with other MIR diagnostic based on ionic lines
of high excitation levels which trace the presence of the AGN radiation
field (\cite{Genzela}). These two methods are complementary since
forbidden line emission due to an AGN in high resolution spectra can
be easily identified in a spectrum (see the SWS spectrum of Circinus
in \cite{Moorwood}) even when a starburst is dominating the total MIR
emission.

\section{Conclusions}
Studying the effects of spatial resolution in our MIR spectra
we conclude that:

1) Our MIR diagnostic diagram based on the ISOCAM spatial resolution
($\sim$8'') can detect the presence of AGNs in late type galaxies only
in the local Universe (D\,$<$\,50\,Mpc).
  
2) Some AGNs associated with ultra-luminous ``monsters'', which are
encountered in merging systems and dominating the integrated MIR
emission, can be detected at large distance such as IRAS\,19254-7245
located at 250\,Mpc.

3) Future instruments with higher spatial resolution will provide
better estimates of the AGN/starburst fraction in more distant
galaxies (D\,$>$\,50\,Mpc).

\vfill
\end{document}